\documentclass[12pt]{iopart}
\usepackage{bm}
\usepackage[dvips]{graphicx}
\usepackage[latin1]{inputenc}
\def\u234{^{234}U}
\begin{document}
\title[Monte Carlo Simulation]{ Monte Carlo Simulation to relate primary and final fragments mass and kinetic energy  distribution from  low energy fission of $\u234$ }
\author{M. Montoya$^{1,2}$, J. Rojas$^{1,3}$ and I. Lobato$^{2}$}
\address{$^{1}$ Instituto Peruano de Energ\'{\i}a Nuclear, Av. Canad\'a 1470, Lima 41, Per\'u.}
\ead{jrojas@ipen.gob.pe}
\address{$^{2}$ Facultad de Ciencias, Universidad Nacional de Ingenier\'{\i}a, Av. Tupac Amaru 210, Apartado 31-139, Lima, Per\'u.}
\address{$^{3}$ Facultad de Ciencias F\'{\i}sicas, Universidad Nacional Mayor de San Marcos, Av. Venezuela s/n, Apartado Postal 14-0149, Lima~-~1, Per\'u.}
\begin{abstract}
The kinetic energy   distribution as a function of mass of final fragments ($m$) from low energy  fission of $\u234$, measured with the Lohengrin spectrometer by Belhafaf {\it et al.}, presents  a peak around $m=108$ and another around $m=122$. The authors attribute the first peak to the evaporation of a large number of neutrons around the corresponding mass number; and the second peak to the distribution of the primary fragment kinetic energy.  Nevertheless, the  theoretical calculations related to primary distribution made by Faust {\it et al.} do not result in a peak around $m=122$. In order to clarify this apparent controversy, we have made a numerical experiment in which the masses and the kinetic energy of final fragments are calculated, assuming an initial distribution of the kinetic energy without peaks on the standard deviation as function of fragment mass. As a result we obtain a pronounced peak on the standard deviation of the kinetic energy distribution around $m=109$, a depletion from $m=121$ to $m=129$, and an small peak around $m=122$, which is not as big as the measured by Belhafaf { \it et al}. Our simulation also reproduces the experimental results on the yield of the final mass, the average number of emitted neutrons as a function of the provisional mass (calculated from the values of the final kinetic energy of the complementary fragments) and the average value of fragment kinetic energy as a function of the final mass.  \\

\textsl{Kewwords}: Monte-Carlo; low energy fission; $\u234$; standard deviation. \\
PACS: 21.10.Gv; 25.85.Ec; 24.10.Lx
\end{abstract}
\submitto{\JPA}
\maketitle
\section{Introduction}
\label{intro}
One of the most studied quantities to understand the fission process is the fission fragment mass and kinetic energy distribution, which is very closely related to the topological features in the multi-dimensional potential energy surface of the fissioning system  ~\cite{moller}. Structures on the distribution of mass and kinetic energy may be interpreted by shell effects on  that potential energy,  determined by the Strutinsky prescription and discussed by Dickmann {\it et al.} \cite{dick} and Wilkins {\it et al.}~\cite{wilkins}. 

\begin{figure}{}
\centering
\includegraphics[width=0.6\textwidth]{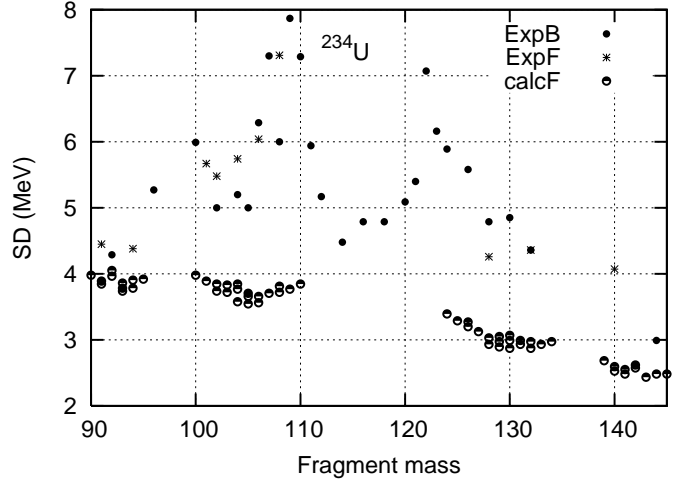}
\caption{ SD of the final fragment kinetic energy distribution as a function of the final mass $m$ ($\bullet$ and $\ast$) as measured by Belhafaf {\it et al.} \cite{belha}, and Faust {\it et al.}  \cite{Faust}, respectively; and SD as a function of primary mass ($\odot$) as calculated by Faust {\it et al.} \cite{Faust}}
\label{fig:sdothers}
\end{figure}

In order to investigate the dynamics of the fission process, the distribution of final fragment kinetic energy ($e$) as a function of final fragment mass ($m$), from thermal neutron induced fission of $^{233}U$, was measured by Belhafaf {\it et al.} \cite{belha}, using the Lohengrin spectrometer. This distribution was represented by the mean value of kinetic energy ($\bar e$) and the standard deviation (SD) $\sigma_{e}$ as  function of $m$. The results present a first peak on $\sigma_e(m)$ around $m=108$ and a second one around $m=122$, see Fig. \ref{fig:sdothers}. The authors attribute the first peak to a large number of evaporated neutrons ($\nu$) around the corresponding primary mass ($A$). Based on the small number of emitted neutron measured around $A=122$, the second peak is attributed to the distribution of the primary fragment kinetic energy ($E$). However, theoretical calculations made by \mbox{Faust {\it et al.} \cite{Faust}} do not result in a peak in SD for the distribution of primary fragment kinetic energy $\sigma_E$ around $A=122$.

In order to clarify this apparent controversy, it is crucial to find the relation between the  primary and the final kinetic energy distributions; the relation between the primary  ($Y(A)$) and the final  ($Y(m)$) mass yield; as well as the relation between the average value of the number of emitted neutron ($\nu = \bar \nu$) as a function of the primary fragment mass and the values corresponding to the experimental results. To address this question, we present Monte-Carlo  simulation results for thermal neutron induced fission of $^{233}U$ i.e low energy fission of  $\u234$ 

\section{Monte Carlo simulation model}
\label{sec:model}
\subsection{Fragment kinetic energy and neutron multiplicity}
In the process of thermal neutron induced fission of $^{233}U$, the excited composed nucleus $^{234}U^*$ is formed first. Then, this nucleus splits in two complementary primary fragments having $A_1$ and $A_2$ as mass numbers, and $E_1$ and $E_2$ as kinetic energies, respectively.
Using relations based on momentum and energy conservation, the total kinetic energy of complementary fragments is,
\begin{equation}
TKE = E_1 + E_2 = \frac{A_1+A_2}{A_1}E_2,
\label{eq:TKE}
\end{equation}
The total excitation energy is given by
\begin{equation}
TXE= Q+\epsilon_n - TKE,
\label{eq:TXE}
\end{equation}
where $Q$ is the difference between fissioning nucleus mass and the sum of two complementary fragments masses, and $\epsilon _n$ is the separation neutron energy of $\u234$. 
From Eq. (\ref{eq:TKE}) and (\ref{eq:TXE}),  taking into account that $A_1 + A_2 = 234$, results
\begin{equation}
TXE = Q + \epsilon_n -\frac{234}{234-A}E,
\label{eq:TXEE}
\end{equation}
where $A$ and $E$ are the primary mass number and kinetic energy, respectively, of one of the two complementary fragments.
It is reasonable to assume that the excitation energy of one complementary fragment ($E^*$) is proportional to the total excitation energy, thus,
\begin{equation}
E^* \propto TXE = Q  + \epsilon_n -\frac{234}{234-A}E,
\label{eq:Ex}
\end{equation}
It is also reasonable to assume that the number ($\nu$) of neutrons emitted by a fragment is proportional to its excitation energy, i.e.
\begin{equation}
 \nu \propto E^*.
\label{eq:nuE}
\end{equation}
From relations (\ref{eq:Ex}) and (\ref{eq:nuE}) we obtain a linear relation between $\nu$ and $E$:
\begin{equation}
 \nu = a + bE,
\label{eq:nuabE}.
\end{equation}
Taking into account that there is no neutron emission ($\nu = 0$) for fragments having the maximal kinetic energy ($E=E_{max}$)  and assuming that $\nu = \bar \nu$ for the average value of fragment kinetic energy, the relation (\ref{eq:nuabE}) becomes:
\begin{equation}
\nu = \bar \nu \left( \frac{E_{max} - E}{E_{max}- \bar E}\right),
\label{eq:nuEmax}
\end{equation}
Let the parameter $\beta$ define the maximal value of kinetic energy by the relation
\begin{equation}
E_{max} = \bar E + \frac{\sigma_E}{\beta},
\label{eq:Emax} 
\end{equation}
Then, the relation (\ref{eq:nuEmax}) may be expressed as
\begin{equation}
\nu = \bar \nu (1 - \beta (\frac{E- \bar E}{\sigma_E})),
\label{eq:nualphsbetaE}
\end{equation}
Because the neutron number $N$ is an integer, it will be defined as the integer part of (\ref{eq:nualphsbetaE}), i.e.
\begin{equation}
N = {\rm Integer ~ part ~ of}(\alpha + \bar \nu (1 - \beta (\frac{E- \bar E}{\sigma_E}))),
\label{eq:nu}
\end{equation}
where $\alpha$ is used to compensate the effect of the change from a real number $\nu$ to an integer number $N$.

\subsection{Simulation process}
In our Monte Carlo simulation the input quantities are the primary fragment yield ($Y$), the average kinetic energy ($\bar E$), the SD of the kinetic energy distribution ($\sigma_E$) and the average number of emitted neutron ($\bar \nu $) as a function of primary fragment mass ($A$). The output of the simulation for the final fragments are the yield ($Y$), the SD of the kinetic energy distribution ($\sigma_e$) and the average number of emitted neutron ($\bar \nu $) as a function of final fragment mass $m$.

For the first simulation, we take $Y$ and $\bar E$ from Ref. \cite{belha}. The first SD $\sigma_E$ curve is an extrapolation of calculation results obtained by  Faust {\it et al.}  \cite{Faust}. Then, we adjust $Y(A)$, $\nu (A)$, $\bar E(A)$ and $\sigma_E(A)$ in order to get $Y(m)$, $\bar \nu $, $\bar e(m)$, $\sigma_e(m)$ in agreement to experimental data.

In the simulation, for each primary mass $A$, the kinetic energy of the fission fragments is chosen randomly from a Gaussian distribution
\begin{equation}
P(E)=\frac{1}{\sqrt{2\pi}\sigma_{E}}
exp\biggl[-\frac{(E-\overline{E})^2}{2\sigma^2_{E}}\biggr],
\label{eq.ETD}
\end{equation}
where $P(E)$ is the probability density of energy with mean value $\overline{E}$  and SD $\sigma_{E}$.

For each $E$ value, the simulated number of neutrons N is calculated with the relation (\ref{eq:nu}). The final mass of the fragment will be, $m = A-N$.
Furthermore, assuming that the fragments loose kinetic energy only by neutron evaporation and not by gamma emission or any other process, and neglecting the recoil effect due to neutron emission,  the kinetic energy $e(m)$ of the final fragment will be given by
\begin{equation}
\label{eq:ef}
e(m)=(1-\frac {N}{A})E,
\end{equation}
With the ensemble of values corresponding to $m$, $e$ and $N$, we calculate $Y(m)$, $\bar e(m)$, $\sigma_e(m)$ and $\nu (m)$.

To obtain an acceptable statistics during the simulation, we have considered a total number of fission events of $\u234$ of the order of $10^{8}$, 
and we have computed the SD of all the relevant quantities by means of the following expression:
\begin{equation}
\sigma^{2}(m)=\frac{\sum_{j=1}^{N_{j}(m)} e^{2}_{j}(m)}{N_{j}(m)}-{\bar e}^2(m),
\label{eq:SD}
\end{equation}
where $\bar e (m)$ is the mean value of the kinetic energy of final fragments with a given mass $m$, and $N_{j}(m)$ is the number of fission events corresponding to that mass.

\section{Results and interpretation}
\label{results}
\begin{figure}
\centering
\includegraphics[angle=270,width=0.5\textwidth]{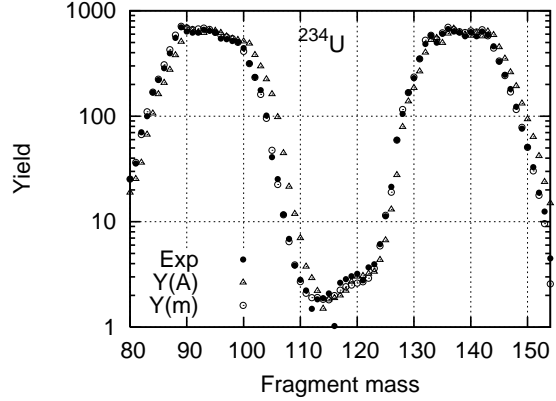}
\caption{ Simulation results for the primary ($\triangle$) and final ($\odot$) mass yields are presented together with experimental data ($\bullet$), taken from Ref. \cite{belha}}.
\label{fig:yield}
\end{figure}

\begin{figure}
\centering
\includegraphics[angle=270,width=0.5\textwidth]{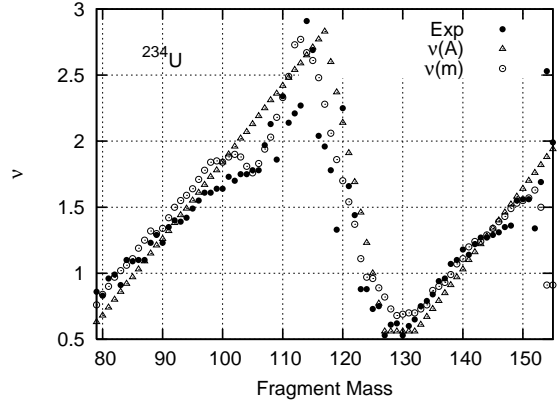}
\caption{The average number of emitted neutrons from fission of $\u234$:~ as a function of the primary fragment mass A ($\triangle$), as a function of final fragment mass ($\odot$) both as result of simulation and experimental data ($\bullet$), taken from Ref. \cite{Nishio}}
\label{fig:neue}
\end{figure}

The simulated final  $Y(m)$ and the primary $ Y(A)$ mass yield curves  are illustrated in Fig.~\ref{fig:yield}. As expected, due to neutron emission, the  $ Y(m)$ curve is shifted from $ Y(A)$ towards smaller fragment masses.  The simulated average number of emitted neutron $\bar \nu (m)$ curve is shifted from $\bar \nu (A)$ in a similar way as $Y(m)$ relative to $Y(A)$(see Fig. \ref{fig:neue}).

As stated in sect.~\ref{sec:model}, the primary kinetic energy ($E(A)$) is generated from a  Gaussian distribution, while the final  kinetic energy ($e(m)$) is calculated through  Eq.~(\ref{eq:ef}).

\begin{figure}
\centering
\includegraphics[width=0.5\textwidth ]{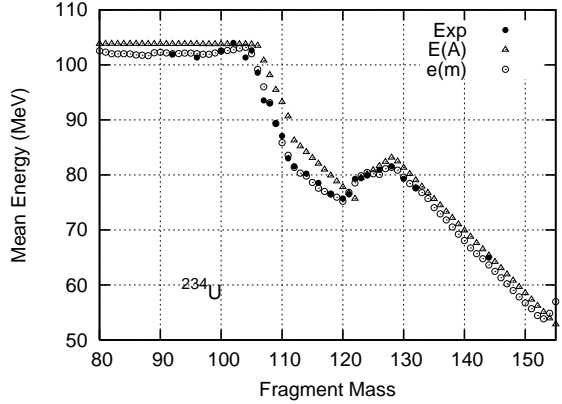}
\caption{ Mean kinetic energy of the primary fragments ($\triangle$)  and mean kinetic energy of the final fragment ($\odot$), as a result of simulation in this work, to be compared to experimental data ($\bullet$)  as measured by Belhafaf {\it et al.}  \cite{belha}.}
\label{fig:ekm}
\end{figure}

The plots of the simulated mean kinetic energy for the primary and final fragments as function of their corresponding masses are shown in Fig.~\ref{fig:ekm}. In general, the simulated average final kinetic energy curve as a function of final mass ($\bar e(m)$) display a shift roughly similar to that of $Y(m)$ curve, and a diminishing of $\bar e$  relative to $\bar E$ values  given by relation (\ref{eq:ef}) with $N=\bar \nu$.

\begin{figure}
\centering
\includegraphics[angle=270,width=0.5\textwidth]{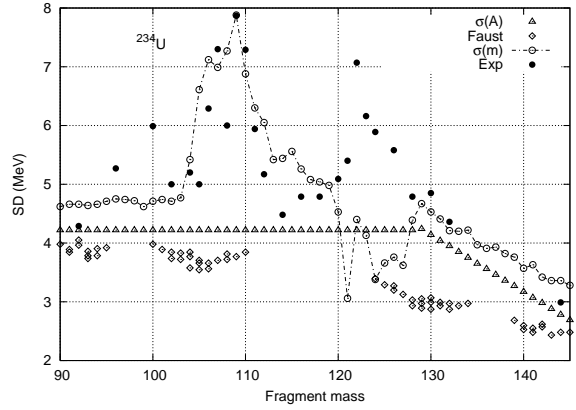}
\caption{Assumed SD of primary fragment kinetic energy distribution  ($\triangle$)  and simulated in this work  final fragment kinetic energy distribution ($\odot$),  to be  compared to results of calculations ($\diamond$) made by Faust {\it et al.}  \cite{Faust} and  experimental data ($\bullet$) as measured by Belhafaf {\it et al.}  \cite{belha} }
\label{fig:sde}
\end{figure}

Furthermore, Fig.~\ref{fig:sde} displays the SD  of the kinetic energy distribution of the primary fragments  ($\sigma_{E}(A)$) and the SD of the kinetic energy of the final fragments ($\sigma_{e}(m)$). The simulated initial distribution of the kinetic energy have no peaks on the SD.  The plots of $\sigma_{e}(m)$ reveal the presence of a  pronounced peak around  $m = 109$ in agreement with the experimental results obtained by Belhafaf {\it et al.}\cite{belha} and Faust {\it et al.} \cite{Faust}, respectively. The peak on the SD around $m=122$ resulting from simulation is not as big as the obtained by Belhafaf et al. Moreover a depletion on the SD in the mass region from $m=121$ to $m=129$ is obtained as a result of simulation.
The  simulated primary fragment kinetic energy distribution (see Fig. \ref{fig:sde},$\triangle$) does not present  peaks in the range of fragment masses $A$ from 90 to 145. If one simulates an additional source of energy dispersion in $\sigma_E$, without any peak, no peak will be observed on $\sigma_e$.
Both the shape and height of the peaks of $\sigma_{e}(m)$ are sensitive to the value of parameter $\alpha$ and $\beta$ appearing in Eq.~(\ref{eq:nu}). A higher value of $\alpha$ will produce a larger peak of SD. The effect of $\beta$ on peak depends much on mass region.  For the region $m=108$, a higher value of $\beta$ will produce a larger peak of $\sigma_e$.  The simulated results for $\sigma_{e}(m)$ presented in Fig. \ref{fig:sde} were obtained with $\alpha$ = 0.62 and $\beta$=0.35. The presence of a peak at $m = 108$ could be associated with neutron emission characteristics (approximately $\bar \nu = 2$) and a very sharp fall in kinetic energy from \mbox{$E$ =96 MeV} to $E$ =90 MeV, corresponding to $A$=109  and $A$=111, respectively. A similar result was obtained for low energy fission of $^{236}U$  \cite{montoya}.
\begin{figure}
\centering
\includegraphics[width=0.5\textwidth]{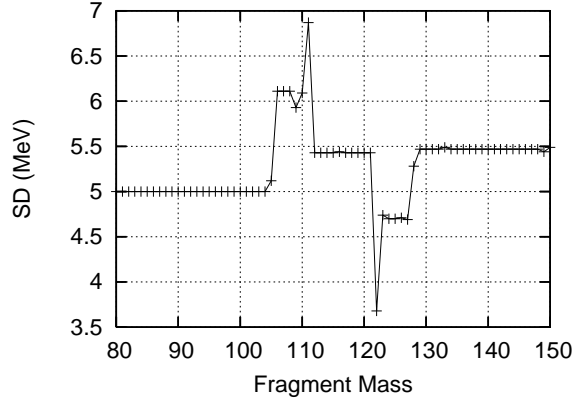}
\caption{SD of final fragments kinetic energy distribution calculated assuming that i) Y(A) is constant, ii) $\sigma _E(A) $ is constant, iii) fragments with $ E > \bar E $ do not emite neutrons and fragments with $ E < \bar E$ emit one neutron and iv) neutron emission have no recoil effect on fragment kinetic energy. $\bar E(A)$ values are taken from data.}
\label{fig:sigmaedeltaE}
\end{figure}

In order to more easily interpret the influence of variation of $\bar E(A) $ on $\sigma_e(m)$, we derive an analytical relation assuming that: i) Y(A) is constant, ii) $\sigma _E(A) $ is constant and iii) fragments with $ E > \bar E $ do not emit neutrons and fragments with $ E < \bar E$ emit one neutron. Then for each final mass there is a contribution from fragments with primary mass $m$ that do not emit any neutron and from fragments with primary mass $m+1$ that emit one neutron. With these conditions we can show that,
\begin{equation}
\sigma_e(m)=\left[\sigma_E^2-\sqrt{\frac{2}{\pi}}\sigma_E \Delta \bar E+\left(\frac{\Delta \bar E}{2} \right)^2\right]^{\frac{1}{2}},
\label{eq:chek}
\end{equation}
where $\Delta \bar E=\bar E(m+1)-\bar E(m)$

The equation (\ref{eq:chek}) corresponds to a parabola with a minimum value

\begin{equation}
\sigma_{e_{min}}= \sqrt{1-\frac{2}{\pi}}\sigma_E = 0.6\sigma_E,
\label{eq:sigmaemin}
\end{equation}

which occurs when

\begin{equation}
\Delta \bar E= 2\sqrt{\frac{2}{\pi}}\sigma_E = 1.6\sigma_E,
\label{eq:sigmaEmin}
\end{equation}

\begin{figure}
\centering
\includegraphics[width=0.5\textwidth]{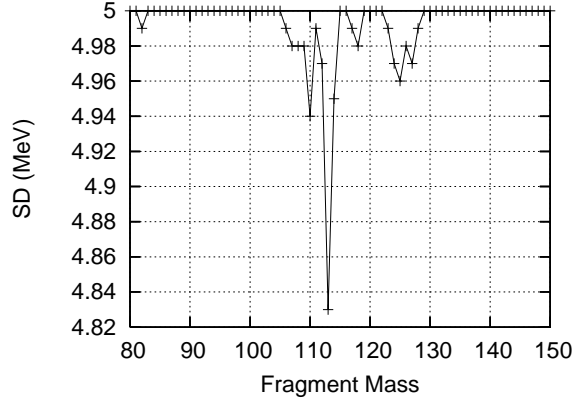}
\caption{SD of final fragments kinetic energy distribution calculated under the assumption thar i) $\sigma _E(A) $ =  5 MeV , iii) $\bar E(m+1)=\bar E(m)$ and iv) neutron emission have no recoil effect on fragment kinetic energy. $Y(A)$ values are taken from data.}
\label{fig:sigmaer}
\end{figure} 
As we can see on Fig.~\ref{fig:sigmaedeltaE}, the $\sigma_e(m)$ curve, calculated with relation (\ref{eq:chek}), presents a peak around $m = 109$ in reasonable agreement with the experimental data.  In that region $\Delta \bar E <0$, then from relation (\ref{eq:chek}) it follows that $\sigma_e(m) >\sigma_E(A)$. The depletion on the simulated $\sigma_e(m)$ on the mass region between $m=121$ and $m=129$ is explained by the fact that $\Delta \bar E > 0$. Using relation (\ref{eq:chek}), we obtain that $\sigma_e(m) <\sigma_E(A)$.

In order to more easily evaluate the influence of the variation of $Y$ on $\sigma_e(m)$, we derive an analytical relation assuming that ( i) $Y(m+1)=r\,Y(m)$, (ii) $\sigma _E(A) $ are constant, (ii) $\bar E(m+1)=\bar E(m)$ and (iii) neutron emission have no recoil effect on fragment kinetic energy. Then we can show that 
\begin{equation}
\sigma_e(m)=\sigma_E\left[1-\frac{2}{\pi}\left(\frac{1-r}{1+r}\right)^2\right]^{\frac{1}{2}},
\label{eq:sigmaeY}
\end{equation}

The $\sigma_e(m)$ values calculated with relation (\ref{eq:sigmaeY}) are lower than $\sigma_E(A)$ and higher than
\begin{equation}
\sigma_e(m)=\sqrt{1-\frac{2}{\pi}}\sigma_E=0.6\sigma_E,
\label{eq:sigmminer}
\end{equation}

The SD curve calculated with this relation is presented in Fig.~\ref{fig:sigmaer}. Using relation (\ref{eq:sigmminer}) we can estimate a peak at $m=122$ assuming that around this mass $Y$ increases very rapidly with $A$ except that  $Y(123) = Y(122)$.  However, we can not reproduce the pronounced peak obtained by Belhafaf {\it et al.}  \cite{belha}. 

\section {Conclusions}
Using a simple model for the neutron emission by fragments, we have carried out a Monte Carlo simulation for the mass and kinetic energy distributions of final fragments from thermal neutron induced fission of $^{233}U$. In comparison with the primary fragments, the final fission fragments have eroded kinetic energy and mass values which give rise to the appearance of peaks in the SD of the final fragments kinetic energy as a function of mass $\sigma_{e}(m)$ around $m$ = 108.  This peak, which agrees with experimental results obtained by Belhafaf {\it et al}. \cite{belha} and Faust {\it  et al.} \cite{Faust} is a consequence of neutron emission and variations of primary fragments yield ($Y(A)$) and mean kinetic energy $\bar E(A)$ curves.  As a result of the simulations one obtains a depletion on the SD in the mass region from $m=121$ to $m=129$, and an small peak on the SD around $m=122$ which is not as big as the measured by Belhafaf {\it et al}.  Our simulation also reproduces the experimental results on the yield of the final mass.

\end{document}